# Electronic energy state and transmission properties study of triple barrier parabolic double quantum well

Mehmet BATI[*]

Recep Tayyip Erdoğan University/Department of Physics, 53100, Rize, Turkey

*mehmet.bati@erdogan.edu.tr

**Abstract.** We investigate transmission properties of triple barrier parabolic double quantum well structure using the non-equilibrium Green's function method. In particular, we examine the effect of system parameters on transmission coefficient. The properties of the electronic state are also studied as a function of the system parameters (such as the well and barrier widths) and electric field bias. We also tested energy electronic state with calculating the density of states. It has been found that the first resonant peaks shift towards the lower energy region as the middle barrier width increases. New resonant peaks emerge when the well widths or depths become wider. These structures are useful for the design of electronic devices.

### 1- Introduction

Over the last decades semiconductor devices have already reached the limit to the nano-scale. Resonant tunnelling properties of semiconductor devices in presence of electric bias firstly have been investigated by Tsu [1, 2]. There are various methods for investigation of tunnelling properties of these devices [3-5]. The non-equilibrium Green's function (NEGF) method is an extensively used method for finding transmission properties in nanoscale devices [3].

Quantum wells are semiconductor structures in which we can observe and control many quantum mechanical effects. Recent developments in nanofabrication technologies have allowed us to fabricate variety types of wells such as parabolic quantum wells (PQWs) which are related to the development of high performance semiconductor devices (infrared detectors, cascade lasers, etc.) [3-11].

The properties of the electronic state in the PQWs were studied experimentally and theoretically [9, 16]. Because of the unique properties (equally spaced electronic spectrum, radiative transitions at the same oscillator frequency etc.), researchers studied the electronic state in the PQWs by using different methods to derive the energy levels [12].

The aim of the present work is to study transmission properties and electronic states in one dimensional double TBPDQW using NEGF method.

### 2- Model and Method

We apply the NEGF method to find electronic Eigen-state and transmission characteristic of triple barrier parabolic double quantum well (TBPDQW). As depicted in Figure 1, $L_i$ (i=1,2…6) denotes the region boundary of the structure.

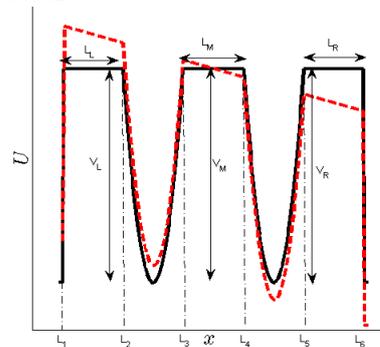

**Figure 1.** Schematic representation of one-dimensional triple barrier parabolic double quantum well, solid line represent without electric field bias (F=0), dashed line under the electric field bias ($F \neq 0$).

One-dimensional PDQW potential can be written as

$$U(x) = \begin{cases} -eFL_1, & L_1 < x \\ -eFx + V_L, & L_1 \leq x < L_2 \\ -eFx + a_L(x - (L_2 + L_3)/2)^2, & L_2 \leq x \leq L_3 \\ -eFx + V_M, & L_3 \leq x \leq L_4 \\ -eFx + a_R(x - (L_4 + L_5)/2)^2, & L_4 \leq x \leq L_5 \\ -eFx + V_R, & L_5 \leq x \leq L_6 \\ -eFL_6, & L_6 < x \end{cases} \quad (1)$$

where, $a_L = \frac{2V_L}{(W_L)^2}$, $a_R = \frac{2V_R}{(W_R)^2}$, left well width is $W_L = |L_3 - L_2|$, and right well width is $W_R = |L_5 - L_4|$ and middle barrier length is $L_M = |L_4 - L_3|$. We use dimensionless form of the Schödinger equation by using the Effective Bohr radius ($a_0^*$) and Hartree energy ($E_H^*$) for scales. So Schrödinger equation becomes

$$-\frac{1}{2}\frac{d^2\psi}{d\tilde{x}^2} + \tilde{U}(\tilde{x})\psi = \tilde{E}\psi \quad (2)$$

Finite difference discretization method applied to the Eq. (2) is as follows [2]:

$$-\tilde{t}\psi_{n-1} + (2\tilde{t} + \tilde{U}_n) - \tilde{t}\psi_{n+1} = E\psi_n \quad (3)$$

where $\tilde{t} = \frac{1}{2\Delta\tilde{x}^2}$ is the hopping parameters, and we use abbreviation $\tilde{U}_n = \tilde{U}(\tilde{x}_n)$. Including self-energies and considering matrix representations, the new form of Eq. (2) becomes

$$[EI - H - \Sigma_L - \Sigma_R]\{\psi\} = \{S\} \quad (4)$$

where [H] is the Hamiltonian matrix, [I] is the identity matrix, $\{\psi\}$ is the wave function vector and $\{S\}$ is scatering term vector. $\Sigma_L$ and $\Sigma_R$ are corresponding to the self-energies of the left and right contacts, respectively. Accordingly, [H] takes form as:

$$\begin{pmatrix} 2\tilde{t} + \tilde{U}_1 & -\tilde{t} & 0 & \cdots & 0 & 0 \\ -\tilde{t} & 2\tilde{t} + \tilde{U}_2 & -\tilde{t} & \cdots & 0 & 0 \\ 0 & -\tilde{t} & 2\tilde{t} + \tilde{U}_3 & \ddots & \vdots & \vdots \\ 0 & 0 & \ddots & \ddots & -\tilde{t} & 0 \\ \vdots & \vdots & \ddots & -\tilde{t} & 2\tilde{t} + \tilde{U}_{N-1} & -\tilde{t} \\ 0 & \cdots & 0 & 0 & -\tilde{t} & 2\tilde{t} + \tilde{U}_N \end{pmatrix} \quad (5)$$

In addition, self-energy terms $[\Sigma_L]$, $[\Sigma_R]$ and source term $\{S\}$ are given by

$$[\Sigma_L] = \begin{pmatrix} -\tilde{t}e^{i\tilde{k}_L\Delta\tilde{x}} & 0 & \cdots & 0 \\ 0 & 0 & & 0 \\ \vdots & & \ddots & \vdots \\ 0 & \cdots & & 0 \end{pmatrix}, [\Sigma_R] = \begin{pmatrix} 0 & \cdots & & 0 \\ \vdots & \ddots & & \vdots \\ 0 & & 0 & 0 \\ 0 & \cdots & 0 & -\tilde{t}e^{i\tilde{k}_R\Delta\tilde{x}} \end{pmatrix}, \{S\} = \begin{pmatrix} -\tilde{t}(e^{i\tilde{k}_L\Delta\tilde{x}} - e^{-i\tilde{k}_L\Delta\tilde{x}}) \\ 0 \\ \vdots \\ 0 \end{pmatrix}.$$

There are several methods for finding the energy eigenvalues corresponding to a particular potential. One can use eigenvalues of [H] matrix to find energy eigenvalues. The retarded Green's function of the system is

$$[G^r] = [(E + i\lambda)I - H - \Sigma_L - \Sigma_R]^{-1} \tag{6}$$

where $\lambda$ is an infinitesimally small positive number. Transmission coefficient T can be computed as

$$T = Tr[\Gamma_L G^r \Gamma_R G^{r+}] \tag{7}$$

Here, $\Gamma_L = i[\Sigma_L - \Sigma_L^+]$ and $\Gamma_R = i[\Sigma_R - \Sigma_R^+]$ are broadening functions. Finally, density of states (DOS) can be computed as follows:

$$DOS = -\frac{1}{\pi} Im(Tr[G^r]). \tag{8}$$

## 3- Results and discussion

Here, transmission coefficient in parabolic quantum well structure is investigated; we assume the effective mass of the electron $0.067m_0$ to be constant through the system. The transmission coefficients are numerically evaluated across TBPDQW structure with different structure parameters. The number of resonance energy states is altering with changing structure parameters. Also, after numeric calculation we convert our result with dimensional form.

We examine the dependence of the transmission coefficient on the well width as depicted in Figure 2. We can see in Figure 2 number of resonance energy levels increases with increasing well width. We see in Figure 2 that increment of well width reduced resonance energy levels. Density of states (DoS) is calculated for 2 nm, 4 nm and 6 nm symmetrical well width structures by calculating the diagonal elements of retarded Green's function. The first peak in the DoS for all structure is sharp and high. Energy level position is consistent with the T(E). First four resonant energy levels reduces with increasing of well width (See figure 2 (c)).

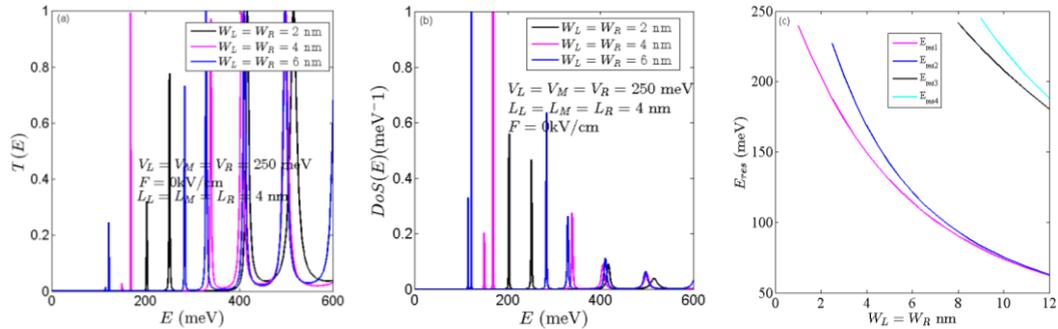

**Figure 2.** (a) Transmission probability and (b) density of states versus electron energy. (c) Changing position of first resonance energy level for different well width. Structure parameter is $F = 0$, $V_L = V_M = V_R = 250$ meV, $L_L = L_M = L_R = 4$ nm.

In order to illustrate the effect of length of middle barrier, we plot Figure 3 for $V_L = V_M = V_R = 250$ meV, $W_L = W_L = 4$ nm, $L_L = L_R = 4$ nm and F=0. We can see in Figure 3 that first resonance energy level decreases but second resonance energy level increases with increasing middle barrier length.

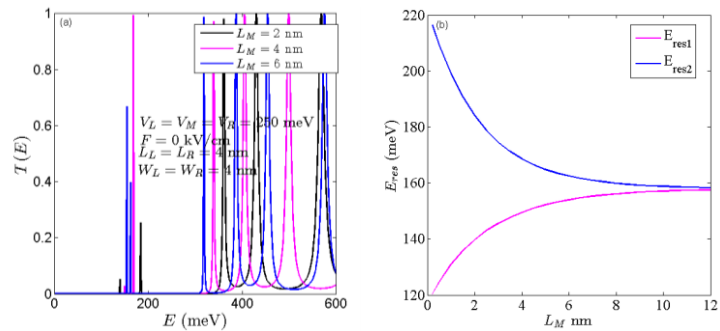
**Figure** 3. (a) Transmission probability versus electron energy and (b) changing position of resonance energy level for different middle barrier widths.

We examine the dependence of the transmission coefficient and position of the resonant energy levels on the electric field bias as depicted in Figure 4. The influence of an electric field bias turns up in a respect of breaking the symmetry of the structure and transmission degreases. From the figure, we see that, first resonance energy level decreases but second resonance energy level increases with increasing electric field bias strength.

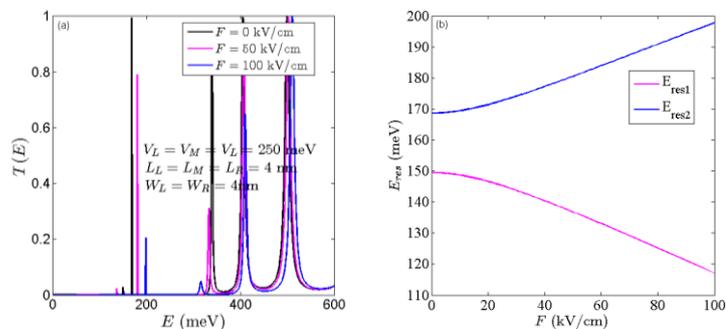
**Figure** 4. (a) Transmission probability versus electron energy for different electric field bias values. (b) Changing position of resonance energy level with electric field bias.

### 4- Conclusions

In summary, resonant tunnelling characteristic for several triple barrier parabolic double quantum well structures are obtained. We have found that structure parameter and electric field bias strongly effected resonant tunnelling properties such as, the first resonant energy states shift towards the lower energy region but second resonant energy states shift towards the higher energy region as the distance between wells increases. These structures are useful for the design of electronic devices.